\newcommand\be{ $$}
\newcommand\ee{ $$ }

\newcommand\bea{\begin{eqnarray}}
\newcommand\eea{\end{eqnarray}}

\documentstyle[twoside,fleqn,espcrc2,epsf]{article}

\newcommand{\AmS}{{\protect\the\textfont2
  A\kern-.1667em\lower.5ex\hbox{M}\kern-.125emS}}

\title{ FROM BLACK HOLES TO GLUEBALLS: \\
The $QCD_3$ Tensor Glueball at Strong Coupling}

\author{Richard  C. Brower\address{Physics Department,
        Boston University, 590 Commonwealth Ave, Boston, MA 02215, USA }
\thanks{ This work was supported in part by the Department
of Energy under Contract No. DE-FG02/19ER40688 and
DE-FG02-91ER40676},
        Samir D. Mathur\address{Center for Theoretical
        Physics, Mass. Inst. of Technology, Cambridge, MA 02139, USA } 
        and Chung-I Tan\address{Physics Department, 
        Brown University, Providence, RI 02912, USA}}

\begin{document}

\begin{abstract}
A strong coupling calculation of glueball masses based on the duality
between supergravity and Yang-Mills theory is presented.  Earlier work
is extended to non-zero spin. Fluctuations in the gravitational metric
lead to the $2^{++}$ state on the leading Pomeron trajectory with a
mass relation: $m(0^{++}) < m(2^{++}) < m(1^{-+})$.  Contrary to
expectation, the mass of our new $0^{++}$ state ($m^2=5.4573$)
associated with the graviton is smaller than the mass of the ${\tilde 0}^{++}$
state ($m^2=11.588$) from the dilaton, which in fact is exactly
degenerate with the tensor $2^{++}$.

\end{abstract}

\maketitle

\section{INTRODUCTION}

The Maldacena conjecture~\cite{maldacena}~ and its further extensions
allow us to compute quantities in a strongly coupled gauge theory from
its dual gravity description. In particular, Witten~\cite{wittenT}~
has pointed out if we compactify the 4-dimensional conformal super
Yang Mills (SYM) to 3 dimensions using anti-periodic boundary
conditions on the fermions, then we break supersymmetry and conformal
invariance and obtain a theory that has interesting mass scales. In
Refs.\cite{csaki} and \cite{jev}, this approach was used to calculate
a discrete mass spectrum for ${\tilde 0}^{++}$ states associated with
$Tr[F^2]$ at strong coupling by solving the dilaton's wave equation in
the corresponding gravity description. Although the theory at strong
coupling is really not pure Yang-Mills, since it has additional
fields, some rough agreement was claimed with the pattern of glueball
masses. Here we report on the calculation of the discrete modes for
the perturbations of the gravitational metric.

\section{\bf AdS/CFT DUALITY AT FINITE T} 

Let us review briefly the proposal for getting a 3-d Yang-Mills theory
dual to supergravity. One begins by considering Type IIB supergravity
in Euclidean 10-dimensional spacetime with the topology 
$M_5\times S^5$.  The  Maldacena conjecture asserts that
IIB superstring theory on $AdS^5\times S^5$ is dual to the
${\cal N } = 4$ SYM conformal field theory on the boundary of the $AdS$ space.
The metric of this spacetime is
\be 
ds^2/R^2_{ads} =  r^2 (d\tau^2+ dx^2_1 + dx^2_2 + dx^2_3) +
{ dr^2 \over r^2 } + d\Omega^2_5 \; , \nonumber
\ee
where the radius of the $AdS$ spacetime is given through $R^4_{AdS} =
g_s N \alpha'^2$ ($g_s$ is the string coupling and $l_s$ is the string
length, $l^2_s = \alpha'$). The Euclidean time is $\tau = i x_0$.  To
break conformal invariance, following \cite{wittenT} , we place the
system at a nonzero temperature described by a periodic Euclidean time
$\tau = \tau + 2 \pi R_0$. The metric correspondingly changes, for
small enough $R_0$, to the non-extremal black hole metric in $AdS$
space. For large black hole temperatures, the stable phase of the
metric corresponds to a black hole with radius large compared to the
$AdS$ curvature scale. To see the physics of discrete modes, we may
take the limit of going close to the horizon,
whereby the metric reduces to that of the black 3-brane. This metric
is (we scale out all dimensionful quantities)
\be
ds^2=f d\tau^2+f^{-1}dr^2+r^2(dx_1^1+dx_2^2+dx_3^2)+
d\Omega_5^2  \; ,
\label{eq:threep} 
\ee 
with
\be
f(r)=r^2-{1\over r^2} \; .
\ee

On the gauge theory side, we would have a $N=4$ susy theory
corresponding to the $AdS$ spacetime, but with the $S^1$
compactification with antiperiodic boundary conditions for the
fermions, supersymmetry is broken and massless scalars are expected to
acquire quantum corrections. Consequently from the view point of a 3-d
theory, the compactification radius acts as an UV cut-off. Before the
compactification the 4-d theory was conformal, and was characterized
by a dimensionless effective coupling $({g_{YM}^{(4)}})^2N\sim g_s
N$. After the compactification the theory is not conformal, and the
radius of the compact circle provides a length scale. Let this radius
be $R_0$. Then a naive dimensional reduction from 4-d Yang-Mills to
3-d Yang-Mills, would give an effective coupling in the 3-d theory
equal to $(g_{YM}^{(3)})^2N=({g_{YM}^{(4)}})^2N/ R_0$. The 3-d YM
coupling has the units of mass. If the dimensionless coupling
$({g_{YM}^{(4)}})^2N$ is much less than unity, then the length scale
associated to this mass is larger than the radius of compactification,
and we may expect the 3-d theory to be a dimensionally reduced version
of the 4-d theory. 

Unfortunately the dual supergravity description applies at
$({g_{YM}^{(4)}})^2N>>1$, so that the higher Kaluza-Klein modes of the
$S^1$ compactification have lower energy than the mass scale set by
the 3-d coupling. Thus we do not really have a 3-d gauge theory with a
finite number of additional fields.  One may nevertheless expect that
some general properties of the dimensionally reduced theory might
survive the strong coupling limit. Moreover, we expect that the
pattern of spin splittings might be a good place to look for
similarities. In keeping with earlier work, we ignore the Kaluza-Klein
modes of the $S^1$ and restrict ourselves to modes that are singlets
of the $SO(6)$, since non-singlets under the $S^1$ and the $SO(6)$ can
have no counterparts in a dimensionally reduced $QCD_3$.

\section{WAVE EQUATIONS}

We wish to consider fluctuations of the metric of the form, 
\be
g_{\mu\nu}=\bar g_{\mu\nu}+ h_{\mu\nu}(x)  \; ,
\ee
leading to the linear Einstein equation,
\be
h_{\mu\nu;\lambda}{}^{\lambda}
+h_\lambda^\lambda{}_{;\mu\nu}
-h_{\mu\lambda;\nu}{}^{\lambda}
-h_{\nu\lambda;\mu}{}^{\lambda} - 8  h_{\mu\nu}=0 \; . \nonumber
\label{eq:thir}
\ee
Our perturbations will have the form
\be
h_{\mu\nu}=\epsilon_{\mu\nu}(r)e^{-mx_3}
\ee
where we have chosen to use $x_3$ as a Euclidean time direction to define 
the glueball masses of the 3-d gauge theory. We fix  the 
gauge to $h_{3\mu}=0$.

From the above ansatz and the metric, we see that we have an $SO(2)$
rotational symmetry in the $x_1-x_2$ space, and we can classify our
perturbations with respect spin.

{\bf Spin-2:} There are two linearly independent perturbations which 
form the spin-2 representation of $SO(2)$:
$h_{12}=h_{21}=q_T(r)e^{-mx_3}  ,  h_{11}=-h_{22}=q_T(r)e^{-mx_3} \;$
with {\rm all ~other~components~ zero}.  The Einstein equations give,
\be
(r^2-{1\over r^2})q_T'' + (r+{3\over r^3})q_T' +
({m^2\over r^2}-4 -{4\over r^4}) q_T = 0. \nonumber
\label{eq:fourt}
\ee
Defining $\phi_T(r)=q_T(r)/ r^2  $,
this is the same equation as that satisfied  by the dilaton
(with constant value on the $S^5$).

{\bf  Spin-1: } The Einstein equation for the ansatz,
$h_{i\tau}=h_{\tau i}=q_V(r)e^{-mx_3}, {i=1 ,2}$
gives 
\be
(r^2-{1\over r^2})q_V'' +(r-{1\over r^3})q_V' +
({m^2\over r^2}-4 +{4\over r^4}) q_V =0 \; .
\ee

{\bf  Spin-0:} Based on the symmetries we choose an ansatz where the 
nonzero components of the perturbation are
\bea
h_{11} &=&h_{22}=q_1(r)e^{-mx_3}\nonumber\\
h_{\tau\tau}&=&-2q_1(r){f(r)\over r^2}e^{-mx_3}+q_2(r)e^{-mx_3}\nonumber\\
h_{rr}&=&q_3(r)e^{-mx_3} \nonumber
\eea
where $f(r)$ is defined above in the metric. The field equation for 
$q_3\equiv q_S(r)$, is
\be
p_2(r) q''_S(r) +   p_1(r) q'_S(r) +  p_0(r)  q_S(r) = 0,
\ee
where $  p_2(r) = r^2(r^4-1)^2 [ 3(r^4-1) + m^2 r^2 ]$, 
$p_1(r) = r(r^4-1)[3(r^4-1)(5r^4+3)+m^2r^2(7r^4+5)]$ and $p_0(r) = 9(r^4-1)^3+2m^2r^2(3+2r^4+3r^8)+m^4r^4(r^4-1)$.

\section{NUMERICAL SOLUTION}

To calculate the discrete spectrum for our three equation, one must 
apply the correct boundary conditions at $r = 1$ and $r = \infty$.
The result is a Sturm-Liouville problem for the propagation of 
gravitational fluctuations in a ``wave guide''.
\begin{table}[hbt]
\setlength{\tabcolsep}{1.0pc}
\newlength{\digitwidth} \settowidth{\digitwidth}{\rm 0}
\catcode`?=\active \def?{\kern\digitwidth}
\caption{Glueball Excitation Spectrum.}
\label{tab:levels}
\begin{tabular}{|l|r|r|r|} \hline
level & $ 0^{++}$ & $ 1^{-+}$  & $ 2^{++}$\\
\hline\hline
 n= 0   & 5.4573   & 18.676 & 11.588 \\ 
\hline
 n= 1   &  30.442  & 47.495 & 34.527 \\ 
\hline
 n= 2   & 65.123   & 87.722   & 68.975 \\ 
\hline
 n= 3   & 111.14     & 139.42  & 114.91 \\ 
\hline
 n= 4   & 168.60    & 203.99 & 172.33 \\ 
\hline
 n= 5   & 237.53   & 277.24 & 241.24 \\ 
\hline
 n= 6   & 317.93   & 363.38 & 321.63 \\ 
\hline
 n= 7   &  409.82  & 461.00  & 413.50 \\ 
\hline
 n= 8   & 513.18    &  570.11 & 516.86 \\ 
\hline
 n= 9   & 628.01   & 690.70 & 631.71 \\
\hline
\end{tabular}
\end{table}
Using this shooting method we have computed the the first 10 states
given in Table \ref{tab:levels}.  The spin-2 equation is equivalent to
the dilaton equation solved in Refs. \cite{csaki} and \cite{jev}, so
the excellent agreement with earlier values validates our method.  
We used a standard Mathematica routine with
boundaries taken to be $x= r^2 -1 = \epsilon$ and $1/x = \epsilon$
reducing $\epsilon$ gradually to $\epsilon = 10^{-6}$.  Note that
since all our eigenfunctions must be even in $r$ with nodes spacing in
$x = r^2-1$ of $O(m^2)$, the variable $1/x$ is a natural way to
measure the distance to the boundary at infinity.  For both
boundaries, the values of $\epsilon$ was varied to demonstrate that
they were near enough to $r = 1, $ and $ \infty$ so as not to
substantially effect the answer.
\begin{figure}[htb]
\epsfxsize=\hsize
\epsffile{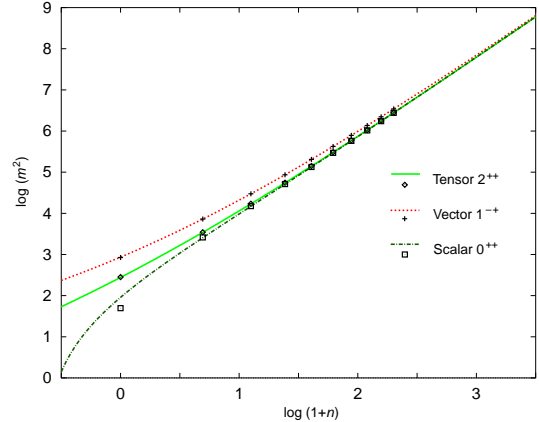}
\vspace{-0.4in}
\caption{WKB fit to Masses in Table 1}
\label{fig:phase}
\vskip -0.5cm
\end{figure}
As one sees in the Fig. \ref{fig:phase}, they match very accurately
with the leading order WKB approximation.
Simple variational forms also lead to very
accurate upper bounds for the ground state ($n = 0$) masses.

\section{DISCUSSION}

Our current exercise needs to be extended to schemes
for 4-d QCD such as the finite temperature versions of $AdS^7 \times
S^4$. As has been suggested elsewhere, the goal may be to find that
background metric that has the phenomenologically best strong coupling
limit. This can then provide an improved framework for efforts to find
the appropriate formulation of a QCD string and for addressing the
question of Pomeron intercept in QCD. In addition, a more
thorough analysis of the complete set of spin-parity states for the
entire bosonic supergravity multiplet and its extension to 4-d
Yang-Mills models is also  worthwhile. Results on
these  computations will be reported in a future publication.

{\noindent\bf Acknowledgments:} We would like to acknowledge useful
conversations with  R. Jaffe, A. Jevicki, D. Lowe, J. M. Maldacena, H.
Ooguri, and others.

\end{document}